\documentclass{emulateapj}

\usepackage{natbib,amsmath}

\newcommand{\kms}{km~s$^{-1}$}
\newcommand{\lya}{Ly$\alpha$}

\def\h2{H$_2$}
\def\f0{$F_0$}

\input epsf

\newcommand{\cm}[1]{\, {\rm cm^{#1}}}

\newcommand{\N}[1]{{N({\rm #1})}}
\newcommand{\sci}[1]{{\rm \; \times \; 10^{#1}}}
\newcommand{\mnhi}{N_{\rm HI}}

\newcommand{\nhi}{$N_{\rm HI}$}

\def\nhi{$N_{\rm HI}$}

\begin{document}

\title{Missing Molecular Hydrogen and the Physical Conditions of GRB Host Galaxies}
\author{Jason Tumlinson\altaffilmark{1}, Jason X. Prochaska\altaffilmark{2}, Hsiao-Wen Chen\altaffilmark{3}, Miroslava Dessauges-Zavadsky\altaffilmark{4},
Joshua S. Bloom\altaffilmark{5}}
\altaffiltext{1}{Yale Center for Astronomy and Astrophysics, Departments
of Physics and Astronomy, Yale University, P. O. Box 208121, New Haven,
CT 06520; jason.tumlinson@yale.edu}
\altaffiltext{2}{University of California Observatories - Lick Observatory, University of California, Santa Cruz, CA 95064; xavier@ucolick.edu}
\altaffiltext{3}{Department of Astronomy and Astrophysics, University of Chicago, 5640 S. Ellis Ave., Chicago, IL 60637; hchen@oddjob.uchicago.edu}
\altaffiltext{4}{Observatoire de Gen\`eve, 51 Ch. des Maillettes,
1290 Sauverny, Switzerland}
\altaffiltext{5}{Department of Astronomy, 601 Campbell Hall, University of California, Berkeley, CA, 94720}
\begin{abstract}

We examine the abundance of molecular hydrogen (\h2) in the spectra of gamma ray burst afterglows (GRBs). In nearby galaxies \h2\ traces the cold neutral medium (CNM) and dense molecular star-forming interstellar gas. Though \h2\ is detected in at least half of all sightlines towards hot stars in the Magellanic Clouds and in $\approx 25\%$
of damped Ly$\alpha$ systems toward quasars, it is not detected in any of the five GRB environments with a similar range of neutral hydrogen column and metallicity. We detect no vibrationally-excited \h2\ that would imply the GRB itself has photodissociated its parent molecular cloud, so such models are ruled out unless the parent cloud was $\lesssim 4$ pc in radius and was fully dissociated prior to the spectroscopic observations, or the star escaped its parent cloud during its main-sequence lifetime. The low molecular fractions for the GRBs are mysterious in light their large column densities of neutral H and expectations based on local analogs, i.e. 30 Doradus in the LMC. This surprising lack of \h2\ in GRB-DLAs indicates that the destruction processes that suppress molecule formation in the LMC and SMC are more effective in the GRB hosts, most probably a combination of low metallicity and an FUV radiation field 10 -- 100 times the Galactic mean field. These inferred conditions place strong constraints on the star forming regions in these early galaxies.
\end{abstract} 

\section{Introduction}

In the Milky Way, stars are born within molecular clouds where cooling by atoms and molecules is sufficient to achieve very high densities. Stars are also observed to form within molecular complexes in the Magellanic clouds and local starbursts exhibit large molecular gas masses. The qualitative connection between molecular gas and star formation is therefore well known in the local universe, although resolving the key physical processes is an area of active research. Theoretical treatments of star formation in the early universe predict that \h2\ is the main coolant of gas with primordial composition \citep{bl04}, but the role of metals and dust in forming the first metal-enriched stars is poorly understood \citep{sch02, ss06}. Current instruments cannot measure directly the molecular mass in galaxies at high redshift, so the metallicity and/or time dependence of star formation has not yet been mapped out in the early Universe.

While we await new mm and radio telescopes that can detect molecular gas at high redshift, absorption-line spectroscopy of quasars and gamma-ray burst (GRB) afterglows offer a means of studying \h2\ in low-metallicity environments, at least in diffuse ISM. Molecular hydrogen has been surveyed extensively in the Milky Way \citep{S77, G06} and Magellanic clouds \citep{h2_mc} by {\em Copernicus} and {\em FUSE}, and in the ISM of high-$z$ galaxies that intercept the sightlines of distant quasars, the damped \lya\ systems \citep[QSO-DLAs;][]{dlas,petit00,ledoux03}.  Because the QSO-DLA population is usually selected according to gas cross-section, quasar sightlines preferentially penetrate the outer regions of a galaxy's ISM and will very rarely intersect molecular clouds with sizes typical of the local universe \citep{zp06}. It is not surprising, then, that one rarely detects gas in DLAs with a high molecular fraction \citep{cui05,pln+06}.

Because their progenitors are massive stars \citep{w93}, long-duration GRBs occur within the star-forming regions of galaxies \citep{bkd02,fls+06}. At UV and optical wavelengths, the unextinguished broadband spectra of GRB afterglows appear consistent with power laws ($f_\nu \sim \nu^{-1} - \nu^{-0.5}$, e.g. \citealt{ksz06}), as expected from a synchrotron shock origin \citep{spn98}. Early-time spectroscopy of these afterglows allows studies of the ISM of the host galaxy and the intergalactic medium along the sightline \citep[e.g.][]{sf04,vel+04,cpb+05}. In this Letter, we survey the \h2\ content of the diffuse gas seen in absorption toward GRB afterglows and compare these against previous measurements of the Milky Way, Magellanic Clouds, and DLA populations. Related papers on the ISM surrounding GRBs describe the metal and dust abundances \citep{pcb+07} and kinematics of the GRB-DLA systems \citep{pcw+07}.

\begin{figure*}[!t]
\centerline{\epsfxsize=\hsize{\epsfbox{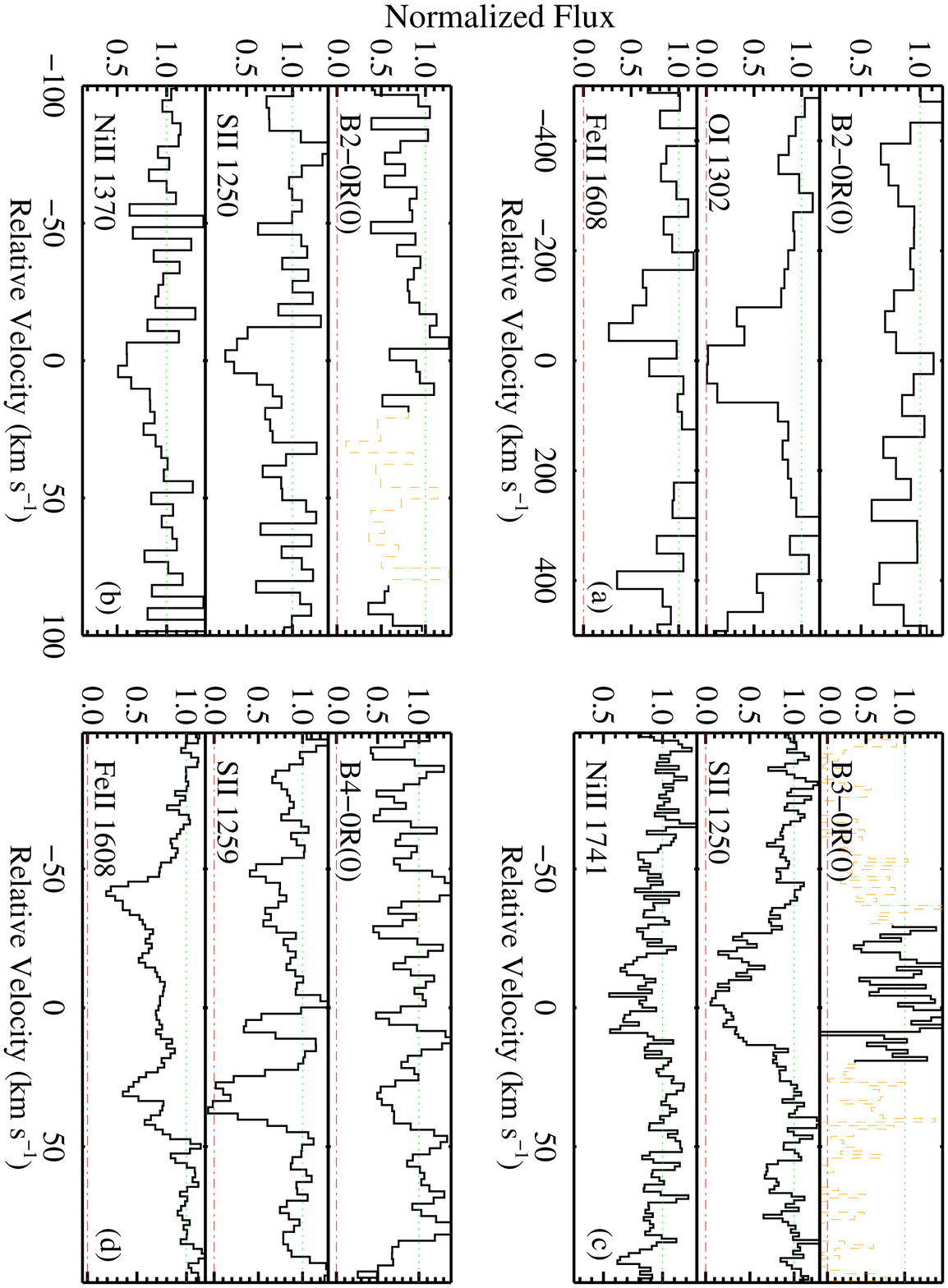}}} \caption{Line-profiles for the strongest undetectable H$_2$ transition and a pair of low-ion transitions which help gauge the metallicity and depletion level of the gas. The \h2\ transitions vary from case to case based on which was least contaminated by extraneous absorption. We show the data for these GRB-DLA: (a) 0303023, (b) 050730, (c) 050820, (d) 050922C. \label{data-fig}} \end{figure*}

\begin{deluxetable*}{cccccccccl}%[!t]
\tablecolumns{3} \tablewidth{0pc} \tablecaption{Data Summary}
\tablehead{GRB & $z_{GRB}$ & $\log{N_{HI}}$ & [M/H]$^a$ & [M/Fe] & Strong Mg$^{b}$& Exc. Fe$^{b}$ & $\log f_{H2}^c$ & $\log N({\rm H}^*_2)^d$ & Ref.}
\startdata
030323  & 3.3720 & 21.90& $>-0.87$ & $>$1.53 & Y & N &  $<-6.5$ & $<13.9$ & 1 \\
050730  & 3.9686 & 22.15& $-2.26$  & 0.25    & ? & Y & $<-7.1$ & $<13.6$ & 2, 3\\
050820  & 2.6147 & 21.00& $-0.63$  & 0.97    & N & N & $<-6.5$ & $<12.9$ & 3\\
050922C & 2.1990 & 21.60& $-2.03$  & 0.75    & W & Y & $<-6.8$ & $<13.5$ & 4\\
060206  & 4.0480 & 20.85& $-0.85$  & \nodata & ? & ? & $<-3.6$ & \nodata & 5
\enddata
\tablenotetext{a}{Metallicity derived from Si, S, or Zn abundance
\citep[see][]{pcb+07}.}
\tablenotetext{b}{See \cite{pcb06}.}
\tablenotetext{c}{With the exception of 060206, the values represent
$4\sigma$ statistical upper limits.}
\tablenotetext{d}{Upper limit ($4 \sigma$) based on non-detection of either L0-3P(1) at 1276.82 \AA\ or L0-3R(2) at 1276.33 \AA\ \citep[see][]{draine02}.}
\tablerefs{
1: \cite{vel+04};
2: \cite{cpb+05};
3: \cite{pcb+07};
4: \cite{pwf+07};
5: \cite{fynbo060206}}
\label{data-table}
\end{deluxetable*}

\section{Data}

The \h2\ molecule exhibits bands of absorption in two electronic transitions (Lyman and Werner; hereafter LW) at rest wavelengths $\lambda = 900 - 1120$\AA\ which are broken into hundreds of individual lines by quantized vibrational and rotational excitation modes (see \citealt{S+B82} for a review of the molecular physics). Approximately 11\% of photoabsorptions in these bands are followed by decay to the vibrational continuum and dissociation of the molecule. Because it forms on the surfaces of dust grains the formation rate of \h2\ is directly proportional to the local dust-to-gas ratio. The balance of these formation and destruction processes determines the molecular fraction of \h2,
$f_{H2} = 2\N{H_2} / [ \N{HI} + 2 \N{H_2}]$. Thus a measurement of $f_{H2}$ and relative populations in the various rotational levels of the ground vibrational state can serve as sensitive diagnostics of local physical conditions such as density, temperature, and ambient FUV radiation field \citep[T02;][]{bts03,splfs05,noterdame07}. For galaxies at $z > 2$, the LW bands are redshifted into the range of ground-based optical telescopes and can be found in the spectra of a bright background source. High spectral resolution ($R \gtrsim$ 30,000) is desirable because (i) the \h2\ lines have intrinsic widths of one to a few \kms\ and (ii) the transitions are located within the \lya\ forest where line-blending can be severe. Table~\ref{data-table} summarizes the five GRB that have been observed at $z>2$ with adequate spectral resolution and coverage of the LW bands.

All of these GRB sightlines exhibited damped \lya\ profiles at the redshift of the host galaxy (termed a GRB-DLA). The \ion{H}{1} column density \nhi\ was derived from Voigt profile fits to the \lya\ transition; all exceed the nominal DLA cutoff of $\mnhi \ge 2\sci{20} \cm{-2}$.  The gas-phase metallicities [M/H] were measured with low-ion transitions of Si, S or Zn assuming no corrections for differential depletion or ionization \citep{pcb+07}.  We also report estimates of the depletion [M/Fe] normalized to the solar ratio and not corrected for any possible nucleosynthetic variations.
The Ti/Fe ratios in GRB-DLA indicate the gas is depleted \citep{pcb+07} and the intrinsic contribution to [M/Fe] is likely only $\approx 0.3$\,dex if we assume standard values observed in metal-poor stars \citep[e.g.][]{mcw97}.

With the exception of GRB~060206, we either acquired the spectra of the sightlines in Table~\ref{data-table} or retrieved the spectra from the ESO-VLT archive \citep[see][for previous analyses]{vel+04,cpb+05,pcb+07,fynbo060206}. We searched for the strongest \h2\ transitions at the systemic redshift (determined from low-ion transitions redward of the \lya\ forest) and selected regions of otherwise unoccupied continuum to avoid coincident \lya\ forest transitions. In all cases we had good coverage of strong $R(0)$ and R(1) transitions (i.e. transitions out of the $J=0$ or $J=1$ rotational levels of the ground vibrational state, see Figure~\ref{data-fig}). Among the four GRB sightlines studied here, none show detectable \h2\ to very low column density limits (4$\sigma$).

\cite{fynbo060206} reported a tentative detection of \h2\ from the W1-0$R(0), R(1)$ and $Q(1)$ transitions toward GRB~060206. An examination of their Figure~1 reveals that the line identified as $Q(1)$ ($f = 0.0365$) has an unexpectedly high equivalent width compared to the combined strength of the $R(0)$ and $R(1)$ lines ($f = 0.0699$ and $0.0340$, respectively). Even in the unlikely case that R(0) makes no contribution to the bluer profile the R(1)/Q(1) ratio should be 0.9 but appears to be substantially lower. This claimed detection based on only three possible lines from two rotational levels in the strongest Werner band and unconfirmed by other bands would not usually count as a detection by the standards employed by, e.g. \cite{h2_mc} in their \h2\ survey of the LMC and SMC. In that study, three unblended lines from at least two bands were needed for a detection. Therefore, we must allow that the absorption lines fitted by \cite{fynbo060206} are instead coincident \lya\ forest lines.  In the following, we adopt the more stringent criterion for a detection and treat their measurement as an upper limit to $f_{H2}$. We perform some statistical tests below counting 060206 as both a detection and upper limit.

\begin{figure}[!t]
\centerline{\epsfxsize=\hsize{\epsfbox{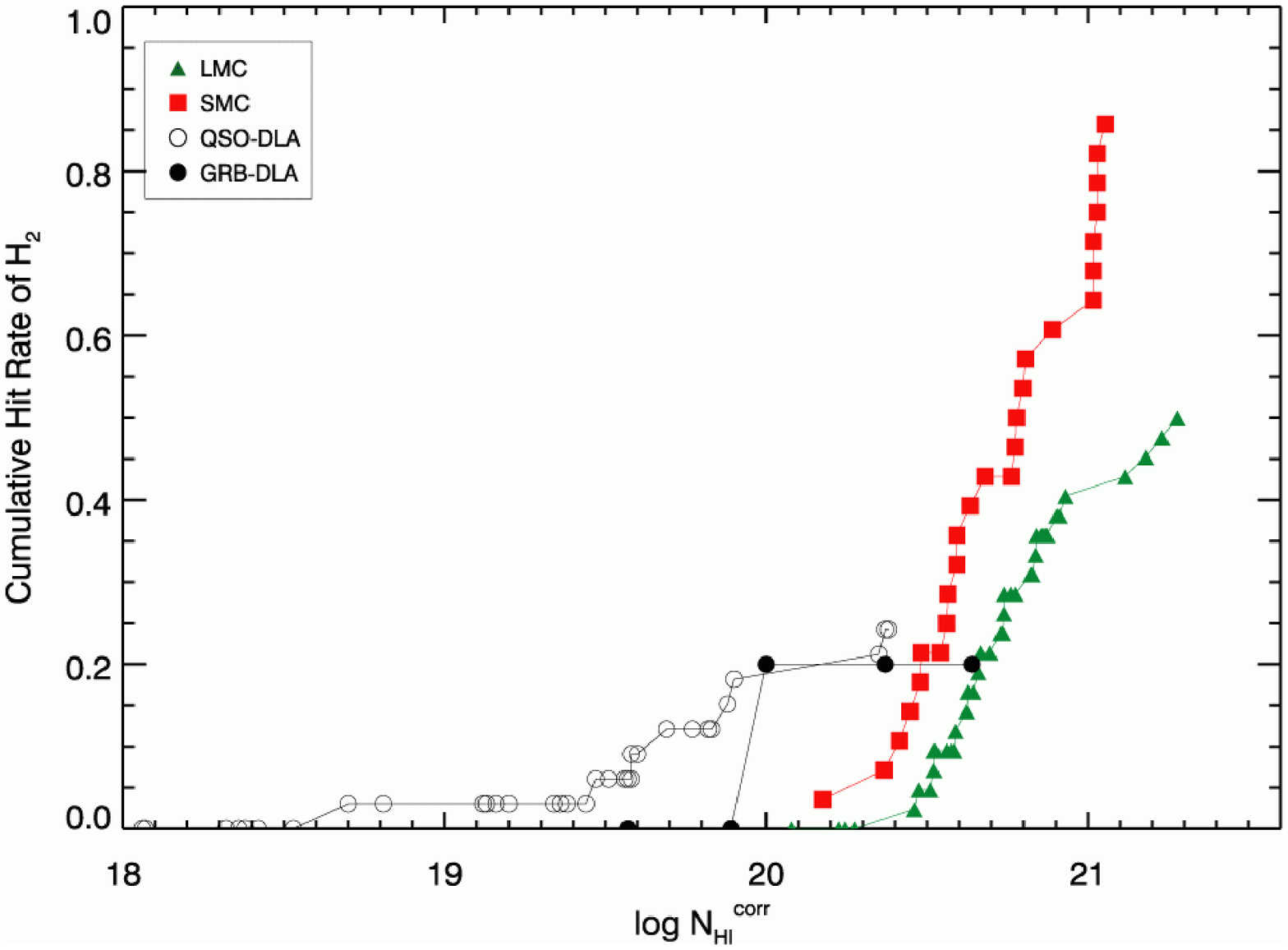}}}
\caption{Cumulative detection rate for the Milky Way (S77), LMC and SMC (T02), and the five GRB afterglows, versus corrected $N_{HI}$. \label{hitrate-fig}} \end{figure}

\section{The Incidence of H$_2$ in GRB-DLA}

The absence of \h2\ in the diffuse gas of GRB-DLAs is somewhat surprising given the large $\mnhi$ associated with these absorbers. Every such sightline in the Milky Way disk studied with {\it Copernicus} \citep{S77} shows \h2. \cite{h2_mc} found molecular hydrogen along 92\% of SMC sightlines and 52\% of LMC sightlines to similar detection limits, over a similar range of $\mnhi$. On the basis of the similar range in $\mnhi$ and [M/H], the QSO-DLAs provide perhaps the best comparison sample for the  GRB-DLAs\footnote{\citet{Hoopes+04} report no detections of \h2\ in FUSE spectra of five star-forming nuclei of nearby galaxies. The interpretation of these results is complicated by the complexity of the background source, since there were many hot stars in the aperture, and $\mnhi$ is difficult to assess. Though that study favors the suppression of \h2\ by intense UV radiation from hot stars, as we do here, we prefer to restrict our comparisons to pencil-beam sightline studies only.}. \cite{ledoux03} found \h2\ in 8 of 33 QSO-DLA systems with $z > 2$ (6 detections in the subset of 24 with $N_{\rm HI} \geq 2 \times 10^{20}$ cm$^{-3}$).

Figure~\ref{hitrate-fig} compares the cumulative detection rate for the MW, LMC, SMC, QSO-DLAs and GRB-DLAs plotted with respect to the ``corrected'' H column, $N_{\rm HI}^{corr} = N_{HI} + {\rm [M/H]}$,
as derived from Table~\ref{data-table}. This corrected H column is designed to place the various environments on a common metal-content scale by normalizing their gas abundances by their gas-to-dust ratios, proxied by [M/H]. This adjustment attempts to remove the dependence of \h2\ fraction on metallicity, leaving radiation field as the remaining major influence on $f_{H2}$. We assume $Z_{LMC} = 0.4$, $Z_{SMC} = 0.2$ and ignore depletion. The GRBs cover roughly the same range of $N_{HI}^{corr}$, but there are no \h2\ detections even to very low limits (discounting GRB060206). Over the $N_{\rm HI}^{corr}$ range, the detection rate is 52\% in the LMC and 92\% in the SMC. The 0 - 20\% detection rate in the GRBs is low, even at the same corrected $N_{HI}$, compared with the Magellanic clouds. The QSO-DLAs resemble the GRB-DLAs in overall detection rate {\em if} the GRB060206 is real, but achieve their maximum fraction at a lower $N_{\rm HI}^{corr}$ than the MCs.

If the GRBs arise in dense star forming regions, perhaps their best local analog of the GRB-DLAs is the 30 Doradus region of the LMC, where hundreds of massive stars have been formed within the last 10 - 20 Myr. Yet in three sightlines toward the edge and center of the region, strong \h2\ absorption with $N$(\h2) $\sim 10^{19} - 10^{20}$ cm$^{-2}$ is seen in all cases with $f_{H2} \sim 0.1$\citep{hartmut01,danforth}. Thus proximity to a massive star forming region does not itself indicate that no diffuse \h2\ will be present, provided the density and metallicity are suitable. This comparison only emphasizes the central mystery of the GRB-DLAs: why do they have such high $N_{\rm HI}$ and such low N(\h2)? We take up this question in the next section.

\section{Why is H$_2$ Absent?}

\subsection{Chance of Statistical Fluctuation?}

Might the 80-100\% percent non-detection rate of GRB-DLAs result from random chance even if the underlying distribution of \h2\ is identical to the other populations? In a sample of 5 with SMC conditions (T02) we should get 1 detection 4.6\% of the time and none 1.0\% of the time. For LMC conditions, we expect 1 detection 19.3\% of the time and none 7.4\% of the time. For the QSO-DLAs \citep{ledoux03}, we find 1 detection 36\% of the time and none 30\% of the time. These differences suggest that the actual detection rate in the GRBs is significantly different from the MCs and possibly rarer than in the QSO-DLAs, although a larger sample is required to rule out the null hypothesis at $99\%$c.l. The available data indicates that diffuse \h2\ is intrinsically less abundant in GRB-DLAs than in environments with similar total gas content and metallicity. This result is unexpected, given that GRBs are associated with massive stars, and that nearby massive stars form in molecular complexes with abundant \h2. We now examine the possible physical causes of the absence of \h2.

\subsection{Absence of Dust?}

One possibility is that \h2\ is absent because dust is missing. \h2\ formation is catalyzed on the surface of dust grains \cite{hs71}, so a dustless (e.g., primordial) cloud will form \h2\ only very slowly in the gas phase. Yet the observed 0.4 to 1.0\,dex depletions of Fe relative to S, Si, and Zn in the GRB-DLAs suggest that dust is present at normal levels given the metallicity \citep{pcb+07}. Even if the dust content, relative to the metallicity, is normal, the dust content relative to the gas content depends on the metallicity itself. The dust-catalyzed formation rate of \h2\ is specified as $Rnn_H$, where R is the formation rate per H atom, empirically estimated at $1 - 3\times10^{-17}$ cm$^3$ s$^{-1}$ in the solar neighborhood \citep{j75}, and $n$ is the total particle density (cm$^{-3}$) and $n_H$ is the number density of neutral H alone. Though the scaling of $R$ with metallicity is expected to follow the overall gas-to-dust ratio, and therefore the metallicity, this scaling has not been tested below the metallicity of the Magellanic Clouds (T02). Thus metallicity may partially account for the reduced detection rate of \h2\ in GRB-DLA, but the effect is present even when the metallicity effect has been removed by $N_{HI}^{corr}$.

\subsection{Destruction by the GRB?}

Another possibility is that molecular gas has been destroyed by the GRB itself. Draine \& Hao (2002; hereafter DH02) calculated the time evolution of molecular gas and dust surrounding a typical GRB afterglow, assuming that the burst occurs in a dense molecular cloud $\sim 5$ pc in radius and with density $n_H = 10^3 - 10^4$ cm$^{-3}$. Their key finding is that even if the cloud starts as fully molecular at the time of the burst, \h2\ molecules and dust will be destroyed efficiently by the advancing soft X-ray and UV radiation. In this case, one may observe a spectrum of vibrationally excited \h2\ lines in the emergent spectrum if the parent cloud is not completely destroyed. A model spectrum of the parent clouds appears in their Figure 9, where strong absorption from \h2\ with $N \sim 10^{21}$ cm$^{-3}$ and numerous lines of vibrationally excited \h2\ are predicted.

\begin{figure*}[!t]
\centerline{\epsfxsize=\hsize{\epsfbox{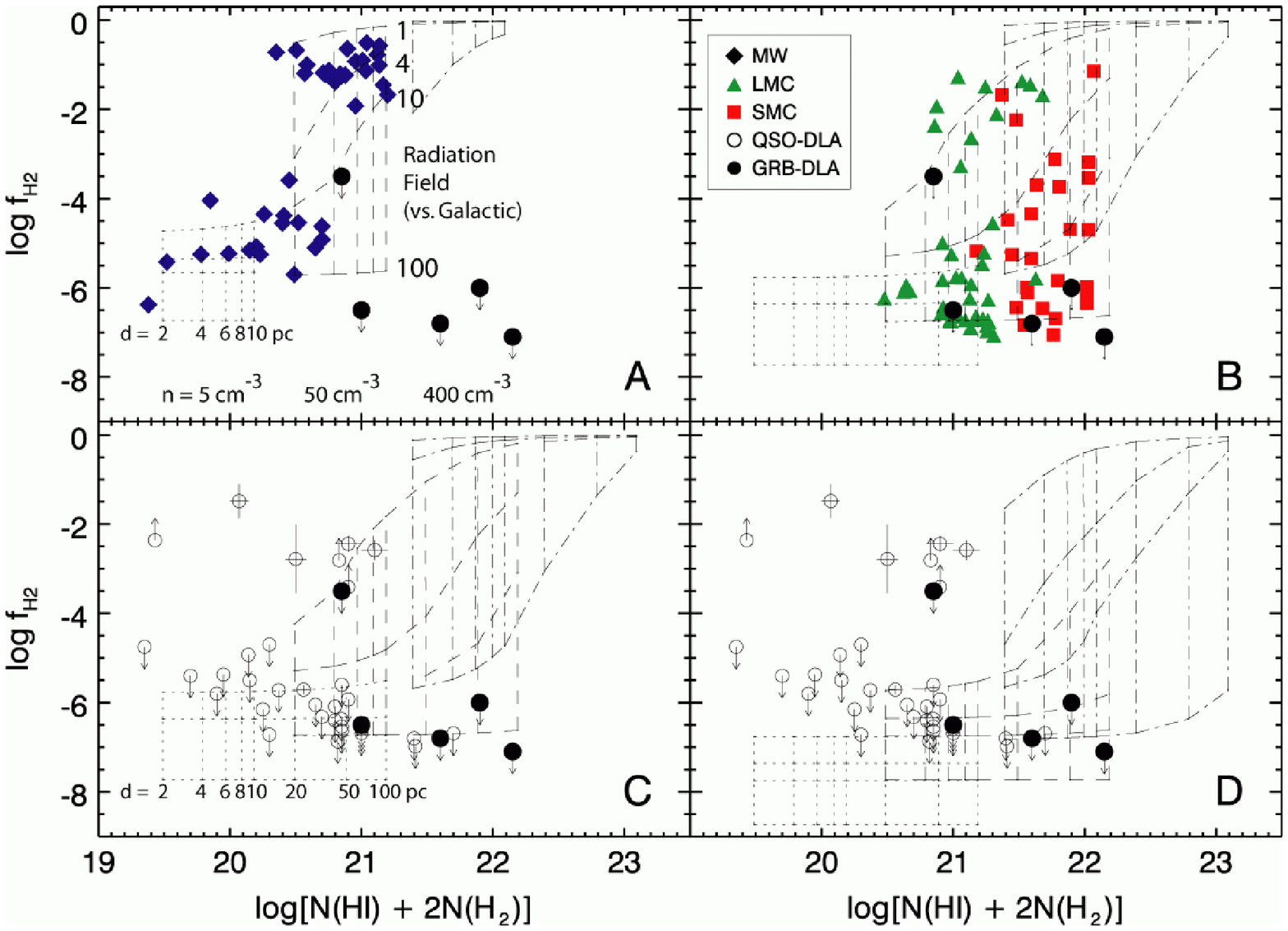}}}
\caption{Molecular fractions compared to models \citep{bts03} for a range of metallicity, cloud size, and FUV radiation field. Three subgrids from left to right in each panel mark cloud densities $n_H =$ 5, 50, and 400 cm$^{-3}$. Panel A: MW points \citep{S77} compared with models with grain formation rate $R = 3 \times 10^{-17}$ cm$^{3}$ s$^{-1}$ for cloud sizes 2, 4, 6, 8, 10 pc (from left to right) and radiation fields 1, 4, 8, and 100 times the Galactic mean. Panel B: LMC/SMC and GRB-DLA molecular fractions compared to 10\% solar metallicity models with cloud sizes 2, 4, 6, 8, 10, 20, 50, and 100 pc and 1, 4, 10, and 100 times the MW radiation field. Panel C: QSO-DLA \citep{ledoux03} and GRB-DLA molecular fractions compared the same model grid as Panel B. Panel D: Same as Panel C but with 1\% solar metallicity models. For statistical purposes, the claimed detection toward GRB060206 at $\log N_{HI} = 20.85$, $f_{H2} = -3.4$ is an upper limit. \label{model-fig}} \vspace{0.1in} \end{figure*}

Neither of these key signatures is seen in any of the GRB-DLA sightlines, to very low detection limits. The strongest predicted \h2\ features (vibrationally excited Lyman 0-3 and 1-2 bands - see Table 1 of DH02), are undetected in our GRB-DLA spectra to limits of $10 - 60$ m\AA. With oscillator strengths $f = 0.03 - 0.05$, these lines have $N \lesssim 1 - 10 \times 10^{13}$ cm$^{-2}$, more than five orders of magnitude smaller than the predictions for clouds undergoing destruction by a GRB\footnote{\cite{fynbo060206} do not report a detection but similar limits on $N({\rm H}_2^*)$ can be estimated by inspection of their Figure 1.}. DH02 state that for a cloud size of 3 pc the ionization/dissociation front reaches the cloud surface $\simeq 20$ s after the burst, after which time the molecular hydrogen has been completely destroyed and leaves no imprint in the spectrum. If the five detected GRBs occurred in dense molecular clouds like those studied by DH02, then clouds of size $\lesssim 4$ pc would have been destroyed by the time the spectra of the afterglow was obtained. Bursts occurring within larger clouds can have their radiation fronts ``stall'', leaving a strong imprint of vibrationally-excited \h2. Thus, the absence of such excitation in our spectra indicate that either the GRBs did not form in and then dissociate their parent clouds, or those clouds were smaller than $\sim 4$ pc in size and so were destroyed before the afterglow was observed.

If GRB host star forming regions resemble local star-forming clouds, it might be unlikely that GRBs ever occur within the parent cloud of the progenitor star. Indeed, in the Milky Way most OB stars are not deeply embedded in their parent molecular clouds, but rather within ionized regions at the boundary of GMC, or in the general ISM hundreds of parsecs from a molecular cloud. Molecular clouds can be driven away from the hot stars formed within them by the ``rocket effect'', in which gas is ionized from the edge of a molecular cloud and the cloud moves away from the star(s) in reaction \citep{1978ppim.book.....S}. The rocket effect can drive clouds at $\sim 10 - 20$ km s$^{-1}$ in the Milky Way disk, transporting them $\sim 50 - 100$ pc away in the $\sim 3 - 5$ Myr lifetime of a massive star \citep{bm90,kmm06}. If this effect is not known to have any explicit dependence on metallicity, so it may act in GRB host galaxies as well. Thus we do not expect that a GRB progenitor would necessarily arise from a star that resides within a dense cloud, even though it must have been formed in one. This possibility justifies treating the GRB-DLA as though the absorbing material were not associated with the GRB progenitor itself.

We thus have no positive evidence for the association of GRBs with dense molecular clouds based on \h2\ observations, even though they lie behind very large HI columns. The absence of \h2\ absorption (as H$_2$ or H$_2^*$) in the afterglow spectra suggests the observed \ion{H}{1} gas is not local to the GRB progenitor. There are additional lines of evidence for placing the \ion{H}{1} gas at distances $\gtrsim 100$\,pc from the GRB progenitor. First, one generally detects the atomic Mg in the GRB afterglow spectrum.  With an ionization potential of $\approx 7.7$\,eV, this atom would be fully ionized by the GRB afterglow if the gas were located within $\approx 100$\,pc of the GRB afterglow \citep{pcb06}. Its detection indicates the majority of gas lies beyond this distance. Second, one finds the X-ray spectrum implies larger columns of metals (O and Fe) than observed in the optical spectrum \citep{watson,butler}. A natural interpretation of this apparent discrepancy is that the gas probed by X-rays is highly ionized and located near the GRB progenitor whereas the neutral gas probed by the optical spectrum lies at much larger distance. For these reasons, we will now proceed by interpreting the molecular fractions ignoring the direct influence of the GRB itself.

\subsection{Constraints on the Physical Conditions of GRB-DLA}

Figure~\ref{model-fig} shows the LMC, SMC, and GRB-DLA molecular fraction patterns compared to model grids calculated with the \cite{bts03} \h2\ radiative transfer code. The models span a range of cloud densities, sizes, and ambient radiation fields, for solar (A), 10\% solar (B and C), and 1\% solar metallicity (D). To explain the reduced abundance of \h2\ in the MCs compared with the MW, T02 invoked both low metallicity and an elevated radiation field owing to the robust star formation. The effect of the UV radiation is clearly seen in Figure~\ref{model-fig}, suggesting that the
apparent lack of \h2\ in the GRBs is also caused by high FUV radiation fields, perhaps from the star-forming region hosting the GRB progenitor. These models show that if we interpret the destruction of \h2\ in the GRB-DLAs as solely caused by ambient FUV radiation fields, radiation fields 10 - 100 times the Galactic mean value are implied. These fields are typical of continuous starbursts of 1 - 10 M$_{\odot}$ yr$^{-1}$ at a distance of $1 - 10$ kpc \citep{s99}. Thus the absence of \h2\ in the GRB-DLAs can be readily explained with the simple hypothesis that the absorbing clouds are illuminated by FUV radiation from recently formed hot stars, as was invoked by T02 for the LMC and SMC. It is not necessary to invoke the direct influence of the GRB UV-optical afterglow itself, though it emits copious amounts of UV radiation \citep{pcb06}. Because it can place upper limits to the amount of radiation incident on the absorbing cloud seen as a GRB-DLA, any detection of \h2\ in such sightlines would be extremely valuable for determining the physical origin of these mysterious absorbers.

%\begin{figure}[!t]
%\centerline{\epsfxsize=\hsize{\epsfbox{0.1solarZ.eps}}}
%\centerline{\epsfxsize=\hsize{\epsfbox{0.01solarZ.eps}}}
%\caption{Upper panel: GRB-DLA and LMC/SMC molecular fractions compared to models \citep{bts03} with a grain formation rate $R %= 3 \times 10^{-18}$ cm$^{3}$ s$^{-1}$, corresponding to 10\% solar metallicity. Results are plotted for semi-infinite slabs %of thickness 2, 4, 6, 8, 10 pc, densities $n_H = $ 5, 50, and 400 cm$^{-3}$, and UV radiation fields 1, 4, 10, and 100 times
%the local Galactic mean. Lower panel: Same as above for $R = 3 \times 10^{-19}$ cm$^{3}$ s$^{-1}$, corresponding to 1\% solar %metallicity \label{bts-model-fig}} \end{figure}

\section{Discussion and Conclusions}

We have studied the molecular hydrogen abundances in five GRB-DLA systems and compared them to diffuse molecular gas in local star-forming environments. These GRB-DLAs have some of the highest HI column densities known, yet show a surprising lack of \h2\ when compared with the Milky Way, Magellanic Clouds, and DLAs toward QSOs. This absence of \h2\ is unexpected given the column densities of \ion{H}{1} and metallicities of these sightlines. We find no evidence that the GRB itself has affected the absorbing material, and in some cases there is positive evidence that it has not \citep{pcb+07}. We therefore conclude that the absorbing material is likely irradiated by a far-ultraviolet radiation field from massive stars that is 10 - 100 times more intense than the interstellar field in the solar neighborhood. This conclusion is broadly concordant with the general conclusion that GRBs arise from massive stars in robustly star forming regions, but the large HI columns and the contrast with analogous local star forming regions (e.g. 30 Doradus) remain to be explained.

In the ISM of nearby galaxies, carbon monoxide (CO) is used as a tracer of star-forming gas in dense environments. CO can be studied in the diffuse interstellar medium using the UV absorption bands from the ground vibrational state. However, detections are challenging in low-column density \h2-bearing clouds even in Galactic material of solar metallicity, which typically show column density ratios N(CO)/N(\h2) $\sim 10^{-7} - 10^{-6}$ for N(\h2) $\simeq 10^{19} - 10^{20}$ cm$^{-3}$ \citep{burgh-co}. Since the abundance of CO should scale down with metallicity, CO absorption from the low-metallicity GRB-DLAs is certainly not expected. Given the easy photodissociation of CO by UV radiation \citep{vdb88}, our result that GRB hosts possess intense FUV radiation fields suggests that CO detections in low-metallicity, star-forming GRB hosts will be extremely challenging. This expectation is corroborated by the recent nondetection of CO emission in the host of GRB030329 \citep{endo07}. We may therefore be left with only indirect far-infrared and mm indicators of dusty star forming gas in high-redshift GRB hosts.

The strong FUV radiation fields inferred for the GRB-DLAs raise the question of whether massive stars are more common in the stellar initial mass function (IMF) in robustly star-forming, low metallicity galaxies at high redshift. There are theoretical reasons to suspect that the characteristic stellar mass increases with decreasing metallicity, owing to reduced cooling efficiency in the gas \citep{omukai05}, or that the IMF shifts to higher mass owing to the thermal influence of radiation in the star-forming environment \citep{rbl05}.  However, the available star-formation indicators for GRB hosts, such as optical nebular emission lines and far-infrared emission from warm dust, trace massive stars only. While some direct or indirect tracer of low and intermediate mass stars must be sought to constrain the IMF more directly, the FUV radiation field inferred for these GRB-DLAs certainly impose a requirement that the IMF must satisfy.

It is difficult to quantify what biases may result from the selection of GRB sightlines by the presence of a bright optical afterglow. To some unknown extent, the GRB-DLAs may represent a special subset of interstellar gas at high redshift. One possible bias is against GRBs that lie behind large column densities of dust, such that they leave no bright optical afterglows that can be studied spectroscopically. Such a system would be likely to show \h2\ as well. But even the systems studied here show (1) large columns of \ion{H}{1}, (2) appreciable dust \citep{pcb+07}, and yet (3) no molecular hydrogen. There is no obvious bias that can select for the first two factors and against the third, so we must interpret these data as if they were fair samples of their ambient ISM. Clearly larger samples are needed to quantify possible selection biases in the GRB ISM sightlines.

\acknowledgements
We acknowledge helpful discussions with Enrico Ramirez-Ruiz, Mark Krumholz, and Bruce Draine. J. T. gratefully acknowledges the generous support of Gilbert and Jaylee Mead for their namesake fellowship in the Yale Center for Astronomy and Astrophysics. J. X. P. is partially supported by NASA/Swift grant NNG05GF55G and an NSF CAREER grant (AST-0548180).

\end{document}